\documentclass[showpacs,aps,twocolumn,prl,floatfix]{revtex4}

\usepackage{amssymb}
\usepackage{amsmath}
\usepackage[dvips]{graphicx}
\usepackage{bm}

\newcommand{\smeq}{\! = \!}

\newcommand{\smpl}{\! + \!}
\newcommand{\smmi}{\! - \!}

\newcommand{\be}{\begin{equation}}
\newcommand{\ee}{\end{equation}}
\newcommand{\bea}{\begin{eqnarray}}
\newcommand{\eea}{\end{eqnarray}}

\begin{document}

\title{Localized Spins on Graphene}
\author{P. S. Cornaglia}
\affiliation{Centro At{\'o}mico Bariloche and Instituto Balseiro,
CNEA, 8400 Bariloche, Argentina}
\affiliation{Consejo Nacional de
Investigaciones Cient\'{\i}ficas y T\'ecnicas (CONICET), Argentina}
\author{Gonzalo Usaj}
\affiliation{Centro At{\'o}mico Bariloche and Instituto Balseiro,
CNEA, 8400 Bariloche, Argentina}
 \affiliation{Consejo Nacional de
Investigaciones Cient\'{\i}ficas y T\'ecnicas (CONICET), Argentina}
\author{C. A. Balseiro}
\affiliation{Centro At{\'o}mico Bariloche and Instituto Balseiro,
CNEA, 8400 Bariloche, Argentina}
 \affiliation{Consejo Nacional de
Investigaciones Cient\'{\i}ficas y T\'ecnicas (CONICET), Argentina}
\date{\today}

\begin{abstract}
The problem of a magnetic impurity, atomic or molecular, absorbed on top
of a carbon atom in otherwise clean graphene is studied using the numerical 
renormalization group. The spectral, thermodynamic, and scattering 
properties of the impurity are described in detail.
In the presence of a small magnetic field, the low energy electronic 
features of graphene make possible to inject spin polarized currents 
through the impurity using a scanning tunneling microscope (STM).
Furthermore, the impurity scattering becomes strongly spin dependent 
and for a finite impurity concentration it leads
to spin polarized bulk currents and a large magnetoresistance. 
In gated graphene the impurity spin is Kondo screened at low temperatures.
However, at temperatures larger than the Kondo temperature, the anomalous
magnetotransport properties are recovered.

\end{abstract}
\maketitle

Graphene is a two dimensional material made of carbon atoms arranged in a
hexagonal lattice. Its structural stability and unusual electronic
properties \cite{Novoselov2004,Novoselov2005,Katsnelson2006,Novoselov2007,Geim2007,CastroNeto-review} make it an excellent candidate for technological
applications. The low energy electronic structure corresponds to
massless, chiral, fermionic quasiparticles described by the Dirac equation. 
Graphene is a semimetal that can be globally or locally doped with electrons or holes
using gate electrodes. These characteristics triggered an intense activity
that ranges from the search of new devices to the study of new scenarios for
Dirac fermions \cite{Geim2007,CastroNeto-review}. Important advances have been made in the preparation and 
characterization of this material. One of the ongoing goals is to
incorporate spintronic effects in graphene and this requires the development of
simple tools for the manipulation and control of the carrier's spins.
There are already some advances in this direction
like the injection of a spin polarized current from ferromagnetic
electrodes \cite{Hugen2008,Tombros2007}. There are also some theoretical proposals involving the use of
edge states to transport spin polarized currents \cite{Wimmer2008,Son2006}. Despite the intense
activity in the area, the properties of graphene with magnetic impurities, 
atoms or molecules, have received less attention. Previous works using mean
field approaches already pointed out the unusual behavior of some
properties as well as the possibility of controlling the magnetic structure
of the impurity with electric fields \cite{Dora2007,Uchoa2008,Sengupta2008}. The magnetic screening
of an impurity spin, the Kondo effect, in systems with graphenelike pseudogaps
has also been analyzed by several authors \cite{Withoff1990,Ingersent1998,Vojta2006}.

In this work we address the problem of graphene with magnetic impurities and
show that the peculiar electronic properties of this system lead to some
interesting new effects. In particular, we show the potential use of these
impurities to inject and generate spin polarized currents. When an impurity
is adsorbed on top of a carbon atom, the impurity levels acquire a finite
lifetime, that is, the spectral function shows broad peaks. In the undoped
case, despite of the broadening of its levels the impurity behaves
as a free spin at low temperatures and we show that with a small magnetic field, 
such that the
Zeeman energy is larger than the thermal energy $k_{B}T$, a \textit{%
non-magnetic} STM tip can be used to inject spin polarized
electrons with an extraordinary efficiency. In this regime we also show
that the bulk transport properties present interesting features: in the absence
of electron--hole symmetry and with a finite impurity concentration there is a large magnetoresistance and the
transport current is spin polarized. Conversely, a magnetic impurity in
doped graphene leads to the Kondo effect.
While in general the
Kondo temperature $T_{K}$ is small due to the small density of states at the
Fermi energy of graphene, in some cases it could be well above the
experimentally accessible limits.

Our starting point is the Anderson model describing an impurity with a
single orbital of energy $\varepsilon_d$ and Coulomb repulsion $U$
hybridized with the conduction electron states with a matrix element 
$V_{\text{hyb}}$. The Hamiltonian of the system is then $H=H_{\text{imp}}+H_{%
\text{graph}}+H_{\text{hyb}}$, where the first term is given by  
\begin{equation}
H_{\text{imp}}=\sum_{\sigma }(\varepsilon_d-\sigma \mu B)d_{\sigma
}^{\dagger }d_{\sigma }^{}+Ud_{\uparrow }^{\dagger }d_{\uparrow }^{}d_{\downarrow
}^{\dagger }d_{\downarrow }^{}.
\end{equation}
Here $d_{\sigma }^{\dagger }$ creates an electron with spin $\sigma $\ at
the impurity state and $\mu B$ is the Zeeman energy shift due to an external
in-plane magnetic field $B$. The Hamiltonian of the graphene layer is 
\begin{equation}
H_{\text{graph}}=-t\sum_{\bm{k},\sigma }\phi (\bm{k})a_{\bm{k}\sigma
}^{\dagger }b_{\bm{k}\sigma }^{}+\phi ^{\ast }(\bm{k})b_{\bm{k}\sigma
}^{\dagger }a_{\bm{k}\sigma }^{},
\end{equation}
where $a_{\bm{k}\sigma }^{\dagger }$ and $b_{\bm{k}\sigma }^{\dagger }$
create electrons with spin $\sigma $ and wavevector $\bm{k}$ on sublattices $A$
and $B$, respectively. The hopping matrix element $t$ is of the order of\ $%
2.7$eV \cite{CastroNeto-review} and $\phi ({\bm k})=\sum_{j}e^{\mathrm{i}\bm{k}\cdot \bm{\delta}_{j}}$
with $\{{\bm\delta }_{j}\}$ the three vectors connecting one site with its 
nearest neighbors. As a result there are two bands of width
$3t$ that touch each other at the corners of the Brillouin zone (Dirac point). 
Finally, assuming that the impurity is adsorbed on a
site of sublattice $A$, the hybridization Hamiltonian is 
$H_{\text{hyb}}=\frac{V_{\text{hyb}}}{\sqrt{N}}\sum_{\bm{k}\sigma }a_{\bm{k}%
\sigma }^{\dagger }d_{\sigma }+d_{\sigma }^{\dagger }a_{\bm{k}\sigma }$,
with $N$ the number of unit cells in the sample.

One of the main ingredients that determines the nature of the solution is the
local density of states of the host material. Close to the Dirac point ($E=0$%
), the local density of states, per atom and spin, can be approximated by $%
\rho (E)\smeq \alpha\left| E\right|$ with $\alpha=A_{uc}/2\pi \hbar ^{2}v_{F}^{2}$, where $A_{uc}$ is
the unit cell area and $v_{F}$ is the Fermi velocity. In what follows we use this form with a high energy cutoff $D$. The other relevant
parameters ($\varepsilon_d$, $U$, $V_{\text{hyb}}$) depend on the nature
of the impurity, in particular $U$ is of the order of a few eVs for
transition metal impurities and smaller for molecules. We solve the
problem using the extensions of Wilson's numerical renormalization group
(NRG) method that allow to describe a non constant density of states for the
host material and to improve the accuracy of the high energy features \cite{Ingersent1998,Bulla2008}. 
We have studied this model numerically for a wide range of parameters. In what follows we
will focus on the localized spin regime where the average number of electrons in the impurity level is 
of the order of one.
\begin{figure}[t]
\begin{center}
\includegraphics[width=0.40\textwidth,clip=true]{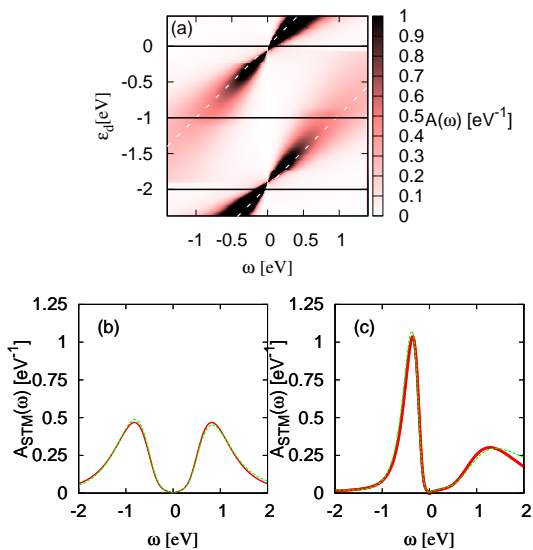}
\end{center}
\caption{(Color online) Color map of the spectral density of a magnetic impurity on graphene. Parameters are 
$U=2$eV, $V_{\textrm{hyb}}=1$eV. (b) STM spectra at the impurity for $t_c/t_d=0$ (solid lines)
and $t_{c}/t_d=0.3$ (dashed lines). Here $\varepsilon_d=-1$eV. (c) Same as
(b) with $\varepsilon_d=-0.5$eV and $t_c/t_d=0,1.0$.}
\label{fig:pd}
\end{figure}

In Fig. \ref{fig:pd}(a) we present a color map of the impurity spectral density $A(\omega)$
for the undoped case-- the Fermi energy $E_F$ laying at the Dirac point. 
The maximums of $A(\omega)$ are shifted from the bare parameters 
$\varepsilon_d$ and $\varepsilon_{d}+U$ shown in the figure with dashed lines.
While for a general $\rho(E)$
some shifts are expected, here the shifts are large and the
energy difference between the peaks is smaller than $U$. This is due to the interpay between a 
Hartree correction and the hybridization self-energy \cite{Uchoa2008}. 
Figs. 1b and 1c show the impurity spectral
functions for two different values of $\varepsilon_d$.
Note the absence of a Kondo peak at $E_F$. 
These spectral
densities could be measured with a STM where electrons from
the microscope tip tunnel to the impurity sensing the local density of
states. If the resonant level is close to the Dirac point ($|\varepsilon
_{d}|\lesssim 1$eV) the STM can measure the resonance. In general
there is some leaking of electrons that tunnel to the substrate generating
Fano structures. The STM differential
conductance at low temperatures is then given by \cite{Cornaglia2003}
\begin{equation}
G(V_{\textrm{b}})=\frac{4e^{2}}{\pi \hbar }\tilde{t}^{2}\rho _{t}A_{STM}(E_{F}+eV_{\textrm{b}}),
\end{equation}
where $V_{\textrm{b}}$ is the voltage drop from the tip to the sample, and $%
\rho _{t}$ is the tip density of states at the Fermi energy. The quantity $A _{STM}(E)$ is the spectral function of
the operator $(t_{c}\psi _{\sigma }^{\dagger }+t_{d}d_{\sigma }^{\dagger })/%
\tilde{t}$, $t_{c}$ and $t_{d}$ are matrix elements for the tunneling of an
electron from the tip to the conduction band states and to the impurity
orbital, respectively, $\tilde{t}=(t_{c}^{2}+t_{d}^{2})^{1/2}$ and $\psi
_{\sigma }^{\dagger }$ is the field operator that creates an electron in a
graphene state centered below the tip. In what follows we consider that the
tip is on top of the impurity and for simplicity we take $\psi _{\sigma
}^{\dagger }=N^{-\frac{1}{2}}\sum_{\bm{k}\sigma }a_{\bm{k}\sigma }^{\dagger }
$. Even for $%
t_{c}\simeq t_{d}$ the effect of $t_{c}$ is very small due
to the smallness of $\rho (E\sim E_{F})$, see Fig. \ref{fig:pd}, 
and the STM gives direct information of the impurity spectral density. 
\begin{figure}[t]
\begin{center}
\includegraphics[width=0.40\textwidth,clip=true]{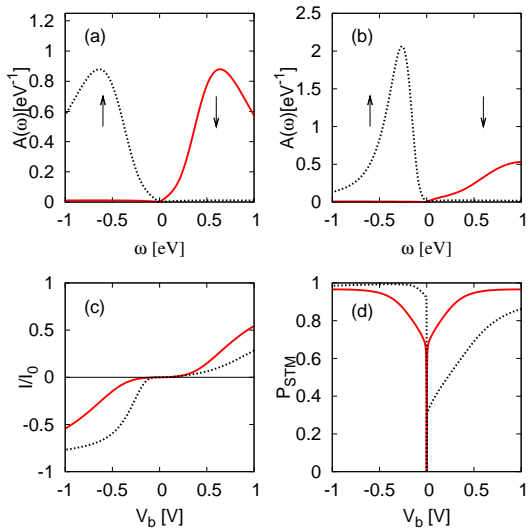}
\end{center}
\caption{(Color online) (a) Impurity spectral density for the undoped case in the presence
of a small magnetic field $\mu B =70 \mu eV$ for the spin up (dashed line) and 
spin down (solid line) projections. Here $V_{hyb}=1.4eV$, $\varepsilon_d=-U/2$ and the other parameters 
are as in Fig. \ref{fig:pd}; 
(b) Same as (a) for $\varepsilon_d=-0.5$eV; (c) STM current and (d) current polarization 
as a function of the bias voltage $V_b$ for the spectral densities 
shown in (a) (solid line) and (b) (dashed line).}
\label{fig:fig2}
\end{figure}

In the presence of an external magnetic field the impurity is polarized and
the spectral densities become spin dependent. In Fig. \ref{fig:fig2} we present results
obtained at low temperatures and low fields. Almost all the weight of the
spin-resolved spectral densities is transferred to a single peak at the
renormalized energies $\widetilde{\varepsilon }_{d}$ (for the spin up) and $%
\widetilde{\varepsilon }_{d}+\widetilde{U}$ (for the spin down). For the
small magnetic field used in the calculation the Zeeman shifts of the peaks are not
appreciable. These results suggest that the spin dependent STM differential
conductance at high voltages becomes very different for the two spin
orientations. To estimate the current of spin-$\sigma $ electrons, $%
I_{\sigma }$, we integrate the spin dependent differential conductance. The
total current $I\smeq I_{\uparrow }\smpl I_{\downarrow }$, in units of $I_{0}\smeq4e%
\tilde{t}^{2}\rho _{t}/\pi \hbar $, and the current polarization $%
P_{STM}\smeq(I_{\uparrow }\smmi I_{\downarrow })/I$ are shown in Figs. \ref{fig:fig2}c and \ref{fig:fig2}d,
respectively. While the tunneling current $I$ is small, the polarization $%
P_{STM}$ can exceed $0.98$.
As we show below the possibility of injecting spin polarized currents is not
restricted to the undoped case.

\begin{figure}[t]
\begin{center}
\includegraphics[width=0.40\textwidth,clip=true]{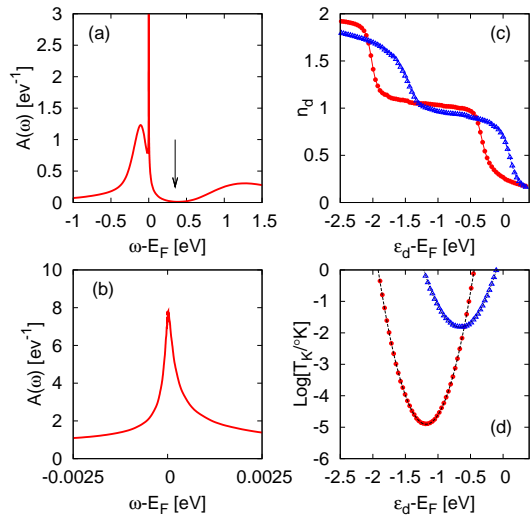}
\end{center}
\caption{(Color online) (a) Impurity spectral density for the doped case ($E_F=-0.35$eV).
The arrow indicates the position of the Dirac point. 
(b) Detail of the Kondo peak. 
(c) Impurity occupation as a function of the level energy; $V_\text{hyb} = 1.4$eV 
and $E_F =-0.35$eV (filled circles); $V_\text{hyb} = 1.75$eV and $E_F = 0.35$eV 
(filled triangles). 
The local interaction is in both cases $U=2$eV. (d) Kondo temperature vs. $\varepsilon_d$
for the parameters of (c). 
The lines are parabolic fits.}
\label{fig:fig3}
\end{figure}

In the case of doped graphene $E_{F}$ is shifted from the Dirac point. In
Fig. \ref{fig:fig3} we present results for the impurity spectral density $A(\omega)$ at
low temperatures. The results show now a Kondo peak at $E_F$.
Following Langreth \cite{Langreth1966}, it can be shown that the spectral
density at $E_{F}$ is given by the usual expression $A(E_{F})\smeq\sin
^{2}(\pi \widetilde{n}_{d}/2)/\pi \Gamma $ where $\widetilde{n}_{d}$ is the
total charge displaced by the impurity and $\Gamma \smeq\pi \rho (E_{F})V_{\text{%
hyb}}^{2}$. Our NRG results reproduce well this exact result. In Fig. \ref{fig:fig3}c we
present results for the impurity charge $n_{d}$
versus $\varepsilon_{d}$. The impurity charge changes when the renormalized
energies $\widetilde{\varepsilon }_{d}$ and $\widetilde{\varepsilon }_{d}+%
\widetilde{U}$ cross $E_{F}$ and the plateau corresponding to a localized spin in the 
impurity, $n_{d}\sim 1$, is narrowed as $V_{\text{%
hyb}}$ increases. This narrowing should not be interpreted as a reduction of
the effective repulsion. As we show below the Kondo temperature is
determined by the bare parameter $U$.

We calculate the impurity spin susceptibility $\chi (T)$, that at low
temperatures shows universal behavior, and extract 
$T_{K}$ using Wilson's criterion $T_{K}\chi (T_{K})/\mu^{2}=0.025$.
The results are shown in Fig. \ref{fig:fig3}d where $\log(T_K)$ as a function 
of $\varepsilon_d$ shows the usual quadratic behavior with a minimum at the 
center of the $n_{d} \sim 1$ plateau and a curvature
determined by the bare interaction $U$. 
For a given set of impurity parameters, $T_{K}$ varies
with doping, or gate voltage, approaching zero for the undoped case. 

\begin{figure}[t]
 \centering
 \includegraphics[bb=50 50 404 404,width=0.35\textwidth]{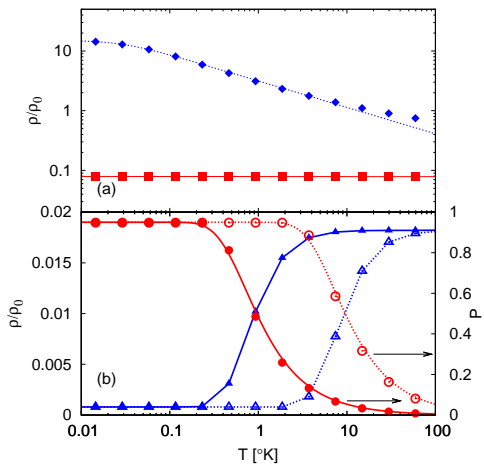}
 \caption{(Color online) (a) Resistivity vs. temperature for
 doped ($E_F=-0.35$eV, solid diamonds) and undoped graphene  
 (solid squares) for $B=0$, $U=2$eV, $V_\textrm{hyb}=1.4$eV, and 
 $\varepsilon_d=-0.63$eV.
 Doped graphene shows Kondo scaling of the resistivity indicated by the dotted line. 
 (b) Resistivity (triangles) and current polarization (circles) vs. temperature.
 Undoped graphene, $\mu B=70\mu$eV (solid symbols) and  $\mu B=700\mu$eV (open symbols).
 Other parameters are $U=3.5$eV and $\varepsilon_d=-0.5$eV.
 }
 \label{fig:4}
\end{figure}

The transport properties of graphene are peculiar in many aspects and it is
interesting to study the effect of adsorbed magnetic impurities. In what
follows we present results for the resistivity $\rho _{\textrm{imp}}(T)$ due to these
impurities in the low impurity concentration ($c_{\textrm{imp}}$) regime. Using the general
expression for the conductivity in graphene \cite{Peres2006,ShonAndo} and evaluating
the band propagators in the Born approximation we obtain, to first order in $c_{\textrm{imp}}$,
\begin{equation}\label{eq:Born}
\rho _{\textrm{imp}}(T)=\rho_0 V_{\text{hyb}}^{2}\left[ \int
 \left( -\frac{\partial f(\omega )}{\partial \omega }\right) \frac{%
\left| \omega \right| }{A(\omega )}\, \textrm{d}\omega\right] ^{-1},
\end{equation}
where $\rho_0 = \pi c_{\textrm{imp}} h/e^{2}$.
The general temperature dependence of the resistivity in the different
regimes requires the numerical evaluation of the integral. As shown
in Fig. \ref{fig:4}, for the undoped case the resistivity is temperature independent while for the doped case we obtain the usual Kondo behavior. 
The NRG results can be qualitatively reproduced performing some simple approximations. 
In the undoped case the spin dependent low frequency
impurity spectral density is given by $A_{\sigma }(\omega )\!\simeq\!
\alpha V_{\text{hyb}}^{2}\left|\omega\right| [(1\smmi n_{d\bar{\sigma} })/\varepsilon
_{d}^{2}\smpl n_{d\bar{\sigma}}/(\varepsilon_d\smpl U)^{2}]$ where $%
n_{d{\sigma} }$ is the number of spin $\sigma $ electrons in the impurity
orbital \cite{Hewson1966}. For the sake of simplicity lets consider first the case $\varepsilon
_{d}=-U/2$ for which  $n_{d\sigma }=\frac{1}{2}$,  the resistivity is then given by 
$\rho _{\textrm{imp}}(T)=\rho_0\alpha V_{\text{hyb}}^{4}/\varepsilon
_{d}^{2}$.
This result corresponds to impurities generating a short range potential of
amplitude $\Delta \propto V_{\text{hyb}}^{2}/|\varepsilon_d|$.  Nevertheless, in
the absence of electron-hole symmetry ($\varepsilon_d\neq -U/2$) the
system presents a large magnetoresistance due to the difference in the
scattering rate of the two spin channels. The contribution of each spin is
given by twice the r.h.s. of Eq. (\ref{eq:Born}) with $A(\omega )$ replaced by $%
A_{\sigma }(\omega )$, and the total resistivity is   
\begin{equation}
\rho _{\textrm{imp}}(T,B)=\rho _{\textrm{imp}}(T,B=0)[1-\gamma ^{2}m^{2}(T,B)],
\end{equation}
with $m(T,B)\smeq(n_{d\uparrow }\smmi n_{d\downarrow })$ and for $n_{d}=1$, $\gamma \smeq((\varepsilon
_{d}\smpl U)^{2}\smmi\varepsilon_d^{2})/((\varepsilon_d\smpl U)^{2}\smpl\varepsilon
_{d}^{2})$. The current polarization is $P\equiv (I_{\uparrow
}\smmi I_{\downarrow })/I\smeq\gamma m(T,B)$. 
These expressions are in good qualitative agreement with the numerical results shown in Fig. 4b.
The actual magnetoresistance and the degree of polarization of the current in
bulk graphene will depend on the presence and intensity of additional scattering 
mechanisms.  For the
magnetoresistance to be observable, the scattering rate due to other
mechanisms should be smaller than the one due to the impurities, that is,
the impurities should be adsorbed on clean graphene.

For the doped case the Kondo screening in the $T\rightarrow0$ limit gives
\begin{equation}
\rho _{\textrm{imp}}(0)=\frac{h}{e^2}\frac{c_{\textrm{imp}}\sin ^{2}(\pi \widetilde{n}_{d}/2)}{\pi^{}n},
\end{equation}
where $n$ is the number of carriers per carbon atom. In the unitary limit, the resistivity
is just determined by the ratio $c_{\textrm{imp}}/n$. In the limit $T\gg T_{K}$ we recover the resistivity characteristic of potential scattering defects.
The temperature dependence of the resistivity, for $T\lesssim T_K$ shows the universal Kondo behavior \cite{Goldhaber-Gordon1998}.

In summary we have numerically solved the problem of a magnetic impurity in graphene 
and analyzed the effect of external in-plane magnetic fields. 
We find that as $\varepsilon_d$ varies, the region of stability for a 
localized spin ($n_{d}\sim1$) is narrowed and shifted with respect to 
the $V_{\textrm{hyb}}\rightarrow0$ limit, in qualitative agreement with mean field results \cite{Uchoa2008}.
Kondo screening of the impurity spin occurs at low temperatures for the doped case. 
The Kondo temperature $T_K$ decreases exponentially with decreasing doping and for the 
undoped case, the impurity behaves as a free spin down to zero temperature. 
We find very little Fano distortions in the STM spectrum that consecuently 
gives a direct measurement of the impurity spectral densities. 
Our central results concern the effect of magnetic fields: for zero or low doping, low $T_K$, the condition $T$,$T_K<\mu B$ is accessible at moderate values of the external field $B$. In this regime the impurity spin is polarized and a non-magnetic 
STM tip can be gated to inject a spin polarized current, that is, the impurity acts as a spin valve. The magnetic field controls the degree and the axis of the spin polarization. In this regime, a finite impurity concentration leads to large magnetotransport effects in bulk graphene: for small values of $|\varepsilon_d-E_F|$ the system shows large magnetoresistance and spin polarized currents. 
All these effects are unique to graphene: they require  $V_{\textrm{hyb}}^2/|\varepsilon _{d}|$ 
to be large and the spin to be free at low temperatures, conditions that are never reached simultaneoulsy in ordinary metals.
Our main results are robust even in the presence of defects or other impurities that 
may change the structure of the pseudogap and the nature of the magnetic screening.
However, for magnetic impurities close to one of these defects, new features
are expected to appear in the STM spectra that will depend on the nature of the defect.

We acknowledge financial support from ANPCyT Grants No 13829/03, No 483/06 and No 482/06 and CONICET PIP 5254/05.  

\bibliographystyle{apsrev}

\end{document}